# Analysis of the User Perception of Chatbots in Education Using A Partial Least Squares Structural Equation Modeling Approach

Md Rabiul Hasan[1], Nahian Ismail Chowdhury[1], Md Hadisur Rahman[1], Md Asif Bin Syed[1], and JuHyeong Ryu[1]

[1]Industrial and Management Systems Engineering, West Virginia University, Morgantown, WV 26505 USA

Corresponding author: Md Rabiul Hasan (e-mail: mh00071@ mix.wvu.edu).

**ABSTRACT** The integration of Artificial Intelligence (AI) into education is a recent development, with chatbots emerging as a noteworthy addition to this transformative landscape. As online learning platforms rapidly advance, students need to adapt swiftly to excel in this dynamic environment. Consequently, understanding the acceptance of chatbots, particularly those employing Large Language Model (LLM) such as Chat Generative Pretrained Transformer (ChatGPT), Google Bard, and other interactive AI technologies, is of paramount importance. However, existing research on chatbots in education has overlooked key behavior-related aspects, such as Optimism, Innovativeness, Discomfort, Insecurity, Transparency, Ethics, Interaction, Engagement, and Accuracy, creating a significant literature gap. To address this gap, this study employs Partial Least Squares Structural Equation Modeling (PLS-SEM) to investigate the determinant of chatbots adoption in education among students, considering the Technology Readiness Index (TRI) and Technology Acceptance Model (TAM). Utilizing a five-point Likert scale for data collection, we gathered a total of 185 responses, which were analyzed using R-Studio software. We established 12 hypotheses to achieve its objectives. The results showed that Optimism and Innovativeness are positively associated with Perceived Ease of Use (PEOU) and Perceived Usefulness (PU). Conversely, Discomfort and Insecurity negatively impact PEOU, with only Insecurity negatively affecting PU. Furthermore, PEOU, PU, Interaction and Engagement (IE), Accuracy, and Responsiveness (AR) all significantly contribute to the Intention to Use (IOE), whereas Transparency and Ethics (TE) have a negative impact on IOE. Finally, IOE mediates the relationships between Interaction, Engagement, Accuracy, Responsiveness, Transparency, Ethics, and Perception of Decision Making (PDM). These findings provide insights for future technology designers, elucidating critical user behavior factors influencing chatbots adoption and utilization in educational contexts.

## I. INTRODUCTION

The advent of artificial intelligence (AI) has transformed human's interactions with technology significantly. AI-driven machines now possess the capability of comprehending human language and responding to inquiries. Chatbots, which simulate human-like conversations and provide personalized assistance, have gained immense popularity. Recent advancement in data set quality, size, and sophisticated techniques for fine-tuning these models with human input have further augmented their capabilities [1]. In the contemporary AI landscape, chatbots also denote a computer program capable of engaging in conversations with users, be it through speech or text [2]. Consequently, these interactive chatbots have gained significant attention in recent years in different areas, including education. Hwang and Chang (2021) have highlighted the advantages of employing chatbots, including enhanced learning efficiency, real-time interaction, and improved peer communication skills [3]. Chocarro et al. (2023) investigated the acceptance of chatbots of teachers in education, employing the Technology Acceptance Model (TAM) [4]. They found that teachers are more inclined to accept chatbots if they perceived them as user-friendly and beneficial. Stathakarou et al. (2020) also investigated students' perceptions of chatbots in healthcare education, discovering that chatbots have the potential to support learning and enhance academic performance [5]. Foroughi et al. (2023) examined the Intention to Use (IOE) of ChatGPT, revealing that performance and effort expectancy, learning value, and hedonic motivation significantly impact its adoption [6]. Chatbots have gained attention not only within classroom learning environments but have also in academic research. Researchers are using machine learning to develop chatbots for education. Følstad and Brandtzæg (2017) pointed out that machine learning and natural language technology are more used among researchers in the development of chatbots for education purposes with the advancement of AI in the field of academic research [7]. Lin et al. (2023) demonstrated that there is a trend to improve the response accuracy of the chatbots in order to make the interaction more interactive, like human conversation [8]. LLM's powered chatbots will lead to a new generation of search engines that can provide detailed and informative answers to complex user questions [9]. In May 2021, Google introduced its LLM, later announced as Google BARD [10]. On the other hand, ChatGPT creates the possibilities of AI-powered educational systems, which have gained exponential interest in recent months. In a recent study, Mogavi et al. (2023) reported that ChatGPT is mostly popular in higher education, while the most discussed topics of ChatGPT are its productivity, efficiency, and ethics [11]. After evaluating the performance of OpenAI, ChatGPT and Google Bard against human ratings, Abdolvahab Khademi (2023) reported that their inter-reliability, as measured by Intraclass correlation (ICC), was low. It indicates that compared to human ratings, considered the gold standard, they did not perform well [12]. Another study conducted by Rahsepar et al. in 2023 compared the accuracy of ChatGPT 3.5, Google Bard, Bing, and Google search engines in responding to a lung cancer questionnaire. The results showed that ChatGPT 3.5 had the highest accuracy rate of 70.8%, compared to Google Bard (51.7%), Bing (61.7%), and Google search engine (55%). However, only some systems achieved 100% consistency [13]. However, it is important to note that this study only focused on the accuracy of responses to a lung cancer questionnaire and did not explore other fields like education or academic search results.

Conversely, the responsible application of AI remains a concern in AI-powered enterprise platforms, including ChatGPT, Google BARD, Notion AI, Jasper, Microsoft Bing, among others. Hoi (2023) argued that various AI principles should be explored, like transparency, reliability and security, ethics, and sustainability [14]. Alsharhan et al. (2023) conducted a literature review on chatbots adoption and demonstrated that the theories that dominate the explanation of chatbots adoption are the technology acceptance model, social presence theory, and the concept of computers as social actors. They also noted that while numerous studies have scrutinized the intention to use chatbots, relatively few have investigated their actual usage and sustained intention [15]. Trust and privacy are another concern when it comes to the word chatbots. Considering it, Lappeman et al. (2023) investigated the trust and digital privacy concerns for banking chatbots services which found that privacy concerns notably impact user self-disclosure, resulting in a negative correlation [16]. This development has increased interests in studying user perceptions of AI-powered chatbots and their future applications. Although, to the best of our knowledge, some previous research work is trying to investigate user behavior towards chatbots and their adoption among the previous research works, there needs to be more research that reflects the user perceptions towards chatbots adoption among students in the academic field in a holistic approach. Our study aims to fill the current gap in research on the behavioral perspectives of chatbots in education. Specifically, we will investigate the accuracy, responsiveness, transparency, ethics, interaction, engagement, and decision-making perception to have overall behavioral perceptions of the students while using chatbots. To assess the effectiveness of LLM-based chatbots among graduate-level students or those considering graduate studies, we will utilize Partial Least Squares Structural Equation Modeling (PLS-SEM). This statistical technique is well-suited to evaluating complex relationships and latent constructs within a model. With PLS-SEM, we can measure the direct effects of these behavior aspects and explore the interplay between them. This approach offers a comprehensive view of chatbots' effectiveness in an educational context. In addition, by using PLS-SEM, we can quantitatively validate and gain a deeper understanding of how these factors influence chatbots' use and perceived impact in education and bridges the gap in the literature and contributes to a more comprehensive understanding of LLM-based chatbots' potential role and improvements in the educational setting.

## II. THEORETICAL BACKGROUND AND HYPOTHESIS

### A. Technology Acceptance Model (TAM)

Many theoretical models help to understand individual behaviors toward new [8]. TAM has been widely adopted as a theoretical framework to understand individual behaviors toward new technology usage. TAM was first introduced by Davis in 1989, which deals with PU, PEOU, and user acceptance of information technology [17]. Previous research has acknowledged the effectiveness of the TAM model and has broadened its scope to encompass external variables crucial for adopting technology [18]. Since the initial introduction of this model, several expansions have been implemented across diverse technologies [19].

### B. Technology Readiness (TR)

Technology Readiness is defined by A. Parasuraman (2000) as people's propensity to embrace and use new technologies for accomplishing goals in home life and at work'. An individual's preparedness to embrace technology can be assessed using the TRI scale, developed by Parasuraman [20].

The follow-up study by Parasuraman and Colby (2015) classified technology adoption into four categories: Innovativeness, Optimism, Discomfort, and Insecurity, which is multi-faceted [21]. Of the four categories, Innovativeness and Optimism have a positive impact, as individuals are more likely to embrace new technology. On the other hand, discomfort, and insecurity act as barriers to TR [22]. According to some researchers, Innovativeness and Optimism are motivating factors that encourage TR, while Optimism and Discomfort act as inhibitors, decreasing an individual's TR [23]. In 2008, Lam et al. conducted an assessment that categorized TR into four dimensions and analyzed the impact of each dimension individually. The assessment highlighted the importance of TR development and its potential outcomes [24]. This study investigates the impact of TR, which includes optimism, innovativeness, discomfort, and insecurity, on TAM, which pertains to PEOU and PU. Different researchers explored the integration of TR and TAM in their research [8][25] [26].

1) OPTIMISM (OP)

Parasuraman (2000) defines optimism as having a positive outlook toward technology and firmly believing that it empowers individuals with greater control, flexibility, and efficiency in their daily lives [20]. People with an optimistic outlook are better equipped to handle negative outcomes, enhancing technology's PEOU and PU (Sohaib et al., 2020) [22]. Based on the information given, someone with an optimistic outlook would view the integration of chatbots as a simple and beneficial process in education. Based on this, we developed two below hypotheses:

Hypothesis 1a (H1a): The level of optimism is directly related to the PEOU regarding the IOE of chatbots in specific fields, such as education.

Hypothesis 1b (H1b): Optimism is positively linked to the PU of chatbots in education.

2) INNOVATIVENESS (IN)

Innovativeness pertains to an individual's readiness to explore novel ideas or methods (Lam et al., 2008). According to Connolly and Kick (2015), innovativeness refers to individuals who enjoy taking risks and finding joy in experimenting with new ideas [27]. Sohaib et al. (2020) found that innovativeness has a clear and positive effect on both PU and PEOU [22]. This supports the idea that being innovative can lead to an increased perception of a product or service's practicality and ease of use. Two hypotheses are proposed:

Hypothesis 2a (H2a): Innovativeness positively correlates with the PEOU towards the IOE of chatbots in education.

Hypothesis 2b (H2b): Innovativeness is positively associated with the PU towards the IOE of chatbots in education.

3) DISCOMFORT (DIS)

It is imperative to acknowledge that technology can induce discomfort when an individual feels inundated by it and senses a lack of authority over its usage (Sohaib et al., 2020). Experiencing discomfort can make it harder to embrace and adopt new technologies, and thus, try to avoid it [20]. Therefore, discomfort has a negative impact on the perceived usefulness (PU) and perceived ease of use (PEOU) of a technology [23]. Two hypotheses are formulated concerning the use of chatbots in education:

Hypothesis 3a (H3a): There is a negative correlation between discomfort and the PEOU of chatbots, affecting the IOE.

Hypothesis 3b (H3b): Discomfort is negatively related to the PU of chatbots, which also impacts the IOE.

4) INSECURITY (INS)

Insecurity pertains to a lack of trust in technology, which arises from doubts about its capability to function correctly and worries about the possible adverse effects it might cause [20]. People with a natural tendency to distrust and be skeptical of technology often assume that there are more risks than benefits associated with it, leading them to avoid using it [23]. Venkatesh et al. (2012) highlighted the importance of trust in influencing individuals' technology adoption behavior [28]. As a result, feeling insecure can harm both PU and PEOU.

The study puts forward two hypotheses. The first one, Hypothesis 4a (H4a), proposes that insecurity negatively impacts the PEOU of chatbots in education, affecting the IOE. The second one, Hypothesis 4b (H4b), suggests that insecurity has a negative impact on the PU of chatbots in education, and this also affects the IOE.

Furthermore, it is crucial to consider the three other hypotheses that pertain to the PEOU, PU, and IOE chatbots in education. Three hypotheses were proposed regarding the use of chatbots in education. The first Hypothesis, H5, suggests that the PEOU of chatbots is positively related to the PU and IOE of chatbots in education. The second Hypothesis, H6, proposes that the PEOU of chatbots is positively related to the IOE of chatbots in education. Finally, the third Hypothesis, H7, suggests that the PU of chatbots is positively related to the IOE of chatbots in education. Researchers have also combined the TAM model with other personalized development approaches, such as TR [8]. Figure 1 below shows the exploration of the TRAM model (TR + TAM) in the first part of this study.

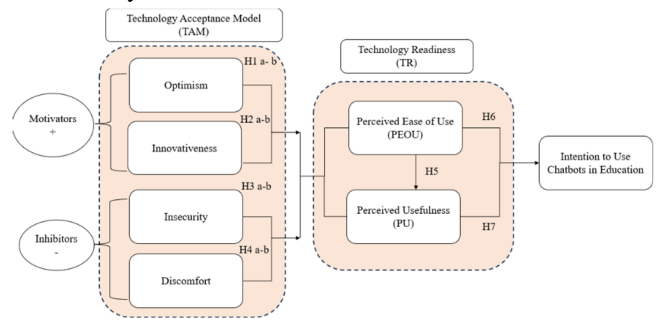

**FIGURE 1.** TRAM model.

Further investigation was conducted to explore how IOE mediates other factors. The study's independent variables include Interaction and Engagement (IE), Accuracy and Speed (AS), and Transparency and Ethics (TE). In contrast, the dependent variable is the PDM with the intention to serve as the mediator, as shown in Figure 2.

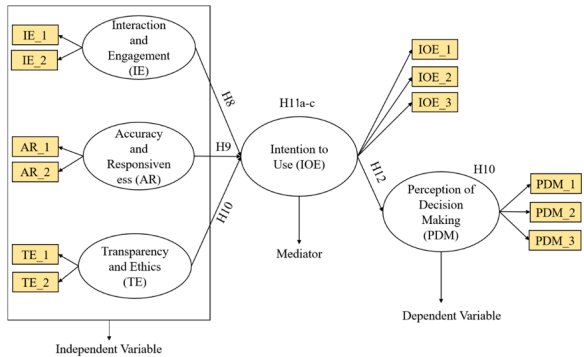

**FIGURE 2.** Conceptual framework and mediating effect

#### 5) INTERACTION AND ENGAGEMENT (IE)
IE is crucial when using chatbots in education. Past research has demonstrated that online interaction and engagement have a significant impact on various positive student outcomes, such as satisfaction and motivation [29] [30] [31]. A positive correlation exists between interaction, engagement, and intention to use.
Hypothesis H8: IE affects the IOE.

#### 6) ACCURACY AND RESPONSIVENESS (AR)
When it comes to chatbots, users expect accurate and timely responses. Kerlyl et al. (2007) emphasized the importance of chatbots learner models in education [32]. Malik et al. (2020) defined responsiveness as the readiness to offer fast help and services. Interestingly, they found no significant association between responsiveness and user intention to use chatbots [33].
Hypothesis H9: A positive correlation between AR and the IOE of chatbots.

#### 7) TRANSPARENCY AND ETHICS (TE)
Chatbots can be opaque, making it hard for users to comprehend their decision-making and answer-generating processes, which raises concerns about transparency [34]. Mozafari et al. (2020) highlighted that disclosing chatbots leads to positive outcomes for certain service types, allowing firms to achieve transparency [35]. Ethics and AI are closely related. Nowadays, chatbots are based on LLM. Virginia Dignum (2018) argued that ethical considerations should be integrated into the design of AI, and ethical reasoning capabilities should be incorporated into the behavior of artificial autonomous systems [36].
Hypothesis H10: TE is correlated with the IOE of chatbots.

#### 8) PERCEPTION OF DECISION-MAKING (PDM)
While conventional chatbots only provide information, chatbots with expert decision-making abilities can solve complex problems. Hsu et al. (2023) discovered that chatbots possessing expert decision-making knowledge significantly improve students' learning achievements [37]. Several studies have explored chatbots' decision-making in various industries, such as healthcare and finance [38][39][40]. The study explores decision-making perspectives with PDM as the dependent variable and IOE as the mediator. The hypotheses being investigated are:
Hypothesis H11a: The IOE mediates the relationship between IE and PDM of education.
Hypothesis H11b: The IOE mediates the relationship between AR and PDM of education.
Hypothesis H11c: The IOE mediates the relationship between TE and PDM of education.
Hypothesis H12: The IOE positively influences the PDM of education.

### III. RESEARCH METHOD
In this study, PLS-SEM was adopted for the study's exploratory nature. The primary appeal of PLS-SEM is that the method enables researchers to estimate complex models with many constructs, indicator variables, and structural paths without imposing distributional assumptions on the data [41]. This method can be adopted for small sample sizes with the absence of distributional assumptions. The method was applied for both analyzing the TRAM model and evaluating the mediation effect, as it has the ability to validate the measurement model and test the structural model hypothesis. SEMinR library in RStudio 2023.06.2+561 was used to perform the analysis. The study received ethical approval from West Virginia University in Morgantown. The protocol number associated with the approval is 2304759788.

#### A. Data collection
A data collection survey was distributed among the participants in the education field. The survey has been circulated among the graduate and undergraduate students from April 2023 through July 2023 for data collection. The survey questionnaire was developed based on the developed hypothesis and distributed among students using Qualtrics. A five-point Likert Scale was used to collect the responses for each question. (5) Strongly agree, (4) Somewhat agree, (3) Neither agree nor disagree, (2) Somewhat disagree, (1) Strongly disagree.

#### B. User Demographics
A total of one hundred and eighty-three people participated in the survey. However, only one hundred and forty-two data were used for analysis as the remaining forty-one responses were incomplete. Among the finalized participants, 58% (82) were male and 42% (60) were female. 35% (50) of participants were in the 18-24 age range and 65% (92) were in the age range of 25-34. On the percentage of Ethnicity among the participants, Asian was 78%, followed by white 14%, Black or African American 4%, and other 10%, respectively. ChatGPT (96%) was the dominant interactive AI or chatbots that participants used, along with interactive AI that ranges from Google BARD, JASPER, Notion AI, Midjourney, and Snapchat AI to Upwork BOT. Most of the participants had a Bachelor's degree, 52% (74), 20% (28) had a Master's degree, 14% (20) were in Ph.D., 4% (5) were doing Post-doc and 11% (15) reported as other degrees. Actual usage frequency was collected in the survey. 27% (38) of participants responded that they use chatbots multiple times a day, 4% (6) of participants use once a day, 26% (37) use few times a week, 3% (4) use once a week, 25% (35) use chatbots few times a month, 9% (13) use once a month and 6% (9) responded their usage frequency as other.

## IV. RESULTS

The assessment of the measurement model involves examining the indicator loading's reliability, internal consistency reliability, convergent validity, and discriminant validity [41]. Indicator loading's reliability has a recommended threshold value of 0.708 or above. The second step is assessing internal consistency reliability; a higher value generally indicates a higher reliability. Jöreskog's (1971) composite reliability (rhoC) is often used for internal consistency reliability, with values ranging from 0.7 to 0.9 are considered satisfactory to good [42]. However, values higher than 0.95 are undesirable since they indicate that the items are redundant, reducing construct validity [43]. Cronbach's alpha is another measure of internal consistency reliability that assumes similar thresholds. However, Cronbach's alpha may be too conservative, and composite reliability may be too liberal. Therefore, Dijkstra and Henseler (2015) proposed rhoA as an improvised measure of construct reliability, which usually lies between Cronbach's alpha and composite reliability [44]. The third criterion evaluates a construct's convergent validity, the average variance extracted (AVE). The minimum acceptable AVE is 0.50 or higher. Hair (2019) recommended the assessment of heterotrait-monotrait ratio (HTMT) of the correlations proposed by Henseler et al. (2015) to evaluate discriminant validity [41] [45]; the fourth criterion to assess measurement model validation. Henseler et al. (2015) propose a threshold value of 0.90 for structural models with constructs that are conceptually very similar, but when constructs are conceptually more distinct, a lower, more conservative, threshold value is suggested, such as 0.85 [45].

### A. Measurement Model Validation

The indicator loading's reliability was over 0.708 for all the latent constructs. Table 1 represents the alpha, rhoC, and rhoA of the variables. It is apparent from this table that all these values are over the recommended value of 0.7, and rhoA lies between alpha and rhoC. In addition, the AVE value of all the variables is over 0.5. Table 2 shows the HTMT values of the measurement model, and all the numbers are less than the recommended value of 0.85. These results validate the TRAM measurement model.

TABLE 1
RELIABILITY AND VALIDITY OF THE MEASUREMENT MODEL

|       | alpha | rhoC  | AVE   | rhoA  |
|-------|-------|-------|-------|-------|
| OP    | 0.814 | 0.877 | 0.642 | 0.824 |
| IN    | 0.705 | 0.870 | 0.771 | 0.722 |
| DIS   | 0.743 | 0.882 | 0.790 | 0.828 |
| INS   | 0.773 | 0.869 | 0.689 | 0.807 |
| PEOU  | 0.779 | 0.900 | 0.818 | 0.799 |
| PU    | 0.834 | 0.923 | 0.857 | 0.837 |
| IOE   | 0.717 | 0.840 | 0.638 | 0.729 |

TABLE 2
DISCRIMINANT VALIDITY (HTMT) OF THE MEASUREMENT MODEL

|      | OP    | IN    | DIS   | INS   | PEOU  | PU    | IOE |
|------|-------|-------|-------|-------|-------|-------|-----|
| OP   | -     | -     | -     | -     | -     | -     | -   |
| IN   | 0.736 | -     | -     | -     | -     | -     | -   |
| DIS  | 0.039 | 0.179 | -     | -     | -     | -     | -   |
| INS  | 0.089 | 0.196 | 0.116 | -     | -     | -     | -   |
| PEOU | 0.563 | 0.411 | 0.138 | 0.152 | -     | -     | -   |
| PU   | 0.776 | 0.693 | 0.083 | 0.101 | 0.503 | -     | -   |
| IOE  | 0.775 | 0.680 | 0.294 | 0.177 | 0.452 | 0.698 | -   |

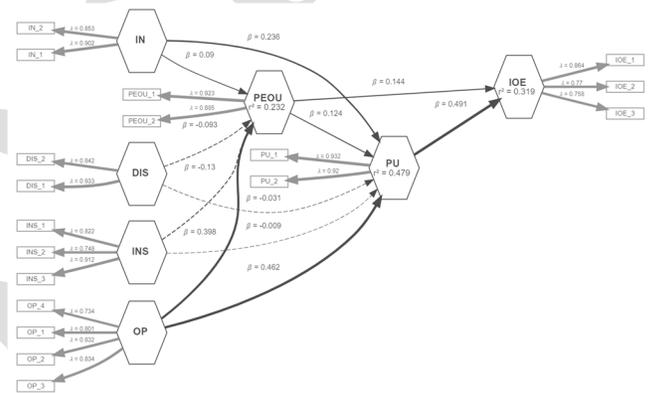

**FIGURE 3.** Structural Model Results

### B. Hypothesis Testing

After validating the measurement model, PLS-SEM was assessed to test the structural model and the hypothesis. The standard assessment criteria for the structural model are the coefficient of determination (R2) [41]. The R2 ranges from 0 to 1, with higher values indicating a greater explanatory power. The recommended values of R2 0.75, 0.5, and 0.25 refer to satisfactory, moderate, and weak, respectively. The results of R2 indicate that 23% (PEOU), 48% (PU), and 32% of the variance is the chatbots use intention (IOE). These values show a weak level of explanation. The path coefficients were evaluated based on the t-test, computed by performing the bootstrapping technique with a significance level of 10%. Bootstrapping is a nonparametric method to test the coefficients, i.e., path coefficients and outer factor weights, by assessing the standard error for estimation. The threshold values for significance levels 10%, 5%, and 1% are 1.65, 1.96 and 2.58 respectively. Figure 3 and Table 3 show the bootstrapped results and the t-test values. According to the results, we fail to reject hypotheses from H1a to H7 except for H3b. From the findings, optimism, and innovativeness have significant effects on the chatbots' perceived ease of use and perceived usefulness, and they positively influence the latent construct. Furthermore, discomfort and insecurity have a negative effect on perceived ease of use, and insecurity has a negative effect on the perceived usefulness of the chatbots, which is indicated by the negative path coefficient value. However, there is not enough evidence to reject hypothesis H3b. Therefore, discomfort does not have a significant effect on perceived usefulness. In addition, perceived ease of use has a positive, significant effect on the perceived usefulness of chatbots in education. Therefore, failing to reject hypothesis H5. Moreover, perceived ease of use and perceived usefulness both have a significant direct effect on the intention to use chatbots in education. Therefore, failing to reject H6 and H7.

TABLE 3
STRUCTURAL MODEL

| Hypothesis | Path | Path Values | Std. | t Value | 5% CI | 95% CI |
|---|---|---|---|---|---|---|
| H1a | OP -> PEOU | 0.398 | 0.095 | 5.947 | 0.409 | 0.719 |
| H1b | OP -> PU | 0.462 | 0.054 | 14.363 | 0.690 | 0.867 |
| H2a | IN -> PEOU | 0.09 | 0.105 | 3.925 | 0.244 | 0.590 |
| H2b | IN -> PU | 0.236 | 0.075 | 9.242 | 0.565 | 0.814 |
| H3a | DIS -> PEOU | -0.093 | 0.084 | 1.651 | 0.061 | 0.331 |
| H3b | DIS -> PU | -0.031 | 0.066 | 1.270 | 0.045 | 0.258 |
| H4a | INS -> PEOU | -0.13 | 0.081 | 1.876 | 0.076 | 0.338 |
| H4b | INS -> PU | -0.009 | 0.043 | 2.338 | 0.078 | 0.219 |
| H5 | PEOU -> PU | 0.124 | 0.108 | 4.679 | 0.321 | 0.675 |
| H6 | PEOU -> IOE | 0.144 | 0.110 | 4.130 | 0.284 | 0.646 |
| H7 | PU -> IOE | 0.491 | 0.076 | 9.219 | 0.573 | 0.821 |

TABLE 4
RELIABILITY AND VALIDITY OF THE MEDIATING MEASUREMENT MODEL

|  | alpha | rhoC | AVE | rhoA |
|---|---|---|---|---|
| IE | 0.713 | 0.874 | 0.776 | 0.721 |
| AR | 0.791 | 0.903 | 0.824 | 0.849 |
| TE | 0.742 | 0.884 | 0.792 | 0.790 |
| IOE | 0.717 | 0.841 | 0.639 | 0.722 |
| PDM | 0.855 | 0.912 | 0.776 | 0.868 |

TABLE 5
DISCRIMINANT VALIDITY (HTMT) OF THE MEDIATING MEASUREMENT MODEL

|  | IE | AR | TE | IOE | PDM |
|---|---|---|---|---|---|
| IE | - | - | - | - | - |
| AR | 0.844 | - | - | - | - |
| TE | 0.109 | 0.214 | - | - | - |
| IOE | 0.778 | 0.688 | 0.284 | - | - |
| PDM | 0.293 | 0.252 | 0.160 | 0.572 | - |

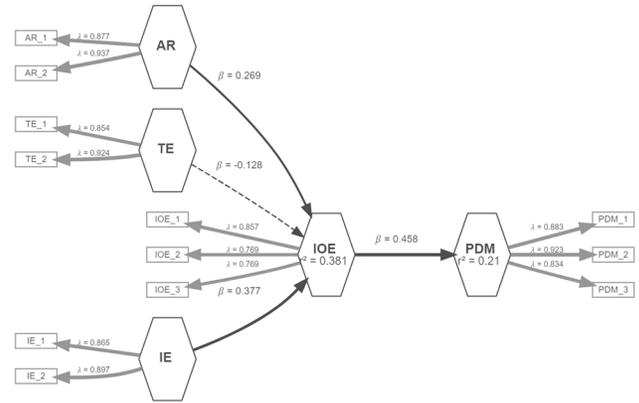

FIGURE 4. Mediating Structural Model Result

The next phase of the research focused on evaluating the user's expectations from the chatbots and how they influence the intention to use and perception of decision-making to use the chatbots. A similar measurement model and structural model were validated using the PLS-SEM. From Tables 4 and 5, it has been found that all the criteria meet or exceed the threshold values, thus validating the measurement model. Furthermore, in a similar way, the structural model was assessed by path coefficients. R2 values show a weak explanatory power of the model with 38% (IOE) and 21% (PDM). Figure 4 and Table 6 show the mediating structural model bootstrapped result. According to the results, this study fails to reject hypotheses H8 and H9. Interaction and engagement (H8), and Accuracy and responsiveness have a direct positive significant effect on intention to use. We fail to reject hypothesis H10 based on a path coefficient that shows Transparency, and ethics have a negative effect on the intention to use chatbots. Lastly, it can be seen from the result that the intention to use has a positive significant effect on the perception of decision-making, and thus rejecting hypothesis H12.

TABLE 6
MEDIATING STRUCTURAL MODEL TESTING

| Hypothesis | Path | Path Values | Std. | t Value | 2.5% CI | 97.5% CI |
|---|---|---|---|---|---|---|
| H8 | IE -> IOE | 0.377 | 0.085 | 9.137 | 0.596 | 0.930 |
| H9 | AR -> IOE | 0.269 | 0.093 | 7.429 | 0.492 | 0.854 |
| H10 | TE -> IOE | -0.128 | 0.115 | 2.468 | 0.108 | 0.554 |
| H12 | IOE -> PDM | 0.458 | 0.108 | 5.295 | 0.353 | 0.774 |

*C. Mediation Effect Analysis*
In this study, the intention to use was determined as a mediator between interaction and engagement (H11a), accuracy and responsiveness (H11b), transparency and ethics (H11c), and perception of decision-making. In the first part of the mediation effect analysis, the indirect effect was significant from IE, AR, and TE to PDM, and IOE was the mediator. Later, direct paths from IE, AR, and TE to PDM were developed, and the bootstrapped model was evaluated to determine the direct effect significance. The results show that the direct effect from IE to PDM is 0.013 with a 95% confidence interval [−0.243; 0.216]. As this interval includes zero, this direct effect is not significant. We, therefore, conclude that IOE fully mediates the relationship between IE and PDM, and we fail to reject H11a. Similarly, the direct effect from AR to PDM is 0.031 with a 95% confidence interval [−0.248; 0.195], and the direct effect is not significant. Therefore, IOE fully mediates the relationship between AR and PDM, and we fail to reject H11b. Finally, the direct effect from TE to PDM is 0.035 with a 95% confidence interval [−0.194; 0.143]. This interval includes zero. Therefore, this direct effect is not significant and concludes that IOE fully mediates the relationship between TE and PDM. We fail to reject H11c.

## V. DISCUSSION

This study aims to determine interactive AI's technological acceptance and readiness in education from a human factor's perspective. The study has shown that technology acceptance behaviors significantly influence user intention to use chatbots. The findings from PLS-SEM analysis show that optimism and innovativeness positively influence the perceived ease of use and perceived usefulness of chatbots. In simple, positive outlook towards the chatbots encourages users to engage with the technology. On the contrary, discomfort, and insecurity related to using chatbots negatively influence the perceived ease of use, and insecurity negatively influences the perceived usefulness of such technology. Therefore, negative traits discourage users from engaging with interactive AI technology. However, it could not be inferred that discomfort has any significant effect on perceived usefulness due to lack of evidence. Moreover, perceived ease of use and perceived usefulness motivate user intention to use interactive AI technology. The study's findings are particularly important to the developers and stakeholders of the technology and draw attention to the user's technology readiness dimensions influencing the use of interactive AI.

The study's findings align with the hypotheses demonstrated by Parasuraman and Colby (2015). According to these authors, optimism and innovativeness are motivators, influencing users to engage more with new technology. However, discomfort and insecurity are inhibitors to accepting and engaging with newer technology; therefore, TRI is a well-defined predictor of technology-related intentional behavior. Furthermore, the study provides insight into other aspects of responsible AI and user concerns in adopting the new technology. The analysis found that interaction and engagement positively influence user intention to use chatbots. In addition, accuracy and responsiveness positively influence users' intention to use chatbots. However, transparency and ethics have a negative effect on the intention to use chatbots. Furthermore, the intention to use fully mediate the decision-making process when using interactive AI chatbots regarding the concerns mentioned above. As mentioned previously, an implication of this study is identifying the user's expectations when using the chatbots and utilizing them to improve the design and interaction of the chatbots technology. The developers and policymakers should design interactive chatbots, considering the users' concerns.

Although our study provides valuable insights, cross-sectional surveys, and self-reported measures may introduce biases. Longitudinal data might be useful to reduce this biasness. Nonetheless, the study's findings still contribute to our understanding of the topic and can serve as a foundation for further research. Future research can be explored on chatbots use by conducting qualitative interviews to assess the user behavior. One study limitation is the small sample size in the data collection. The future study can focus on collecting larger data sets to solidify the conclusion. The study's findings can be compared with other analytical methods like artificial neural networks. Similar studies with different methods can help to shed light on the strengths and weaknesses of the used methods.

## VI. CONCLUSION

The present study was designed to determine the effect of technology readiness dimensions on using and adapting emerging interactive AI technology. This study also contributes to the following findings of chatbots in education: 1. Both discomfort and insecurity have a negative impact on the perceived ease of use of chatbots, but for the perceived usefulness of chatbots, insecurity is negatively related, not discomfort. Therefore, the interactive AI industry should address and mitigate users' discomfort and insecurity towards using interactive AI and increase users' engagement. 2. Perceived ease of use, perceived usefulness, interaction and engagement, accuracy and responsiveness have a significant direct effect on the intention to use while transparency and ethics have negative effect. 3. Intention to use has a positive significant effect on the perception of decision-making; 4. Examining the intention of use as a mediator between interaction and engagement, accuracy and responsiveness, transparency and ethics, and perception of decision-making. Knowing this research model and its outcome will benefit future studies among the researchers, students, and their adoption of chatbots and technology designers.